\documentclass[aps,pre,onecolumn,11pt]{revtex4}
\usepackage[english]{babel}
\usepackage{hyperref}
\usepackage{indentfirst}
\usepackage{graphics}
\usepackage{epstopdf}
\usepackage{epsfig}
\usepackage{float}
\usepackage{ragged2e}
\usepackage{chngcntr}

\newcommand{\ud}{\,\mathrm{d}}

\begin{document}

\title{Collective response to perturbations in a data-driven fish
  school model}
\date{\today}
\author{Daniel S. Calovi$^{1,2}$, Ugo Lopez$^{1,2,3}$, Paul
Schuhmacher$^{1,2}$, Hugues Chat\'e$^4$, Cl\'ement Sire$^{5,6}$, and Guy
Theraulaz$^{1,2}$}

\affiliation{$^1$Centre de Recherches sur la Cognition Animale,
UMR-CNRS 5169, Universit\'e Paul Sabatier, 118 Route de Narbonne,
31062 Toulouse Cedex 4, France}

\affiliation{$^2$CNRS, Centre de Recherches sur la Cognition
Animale, F-31062 Toulouse, France}

\affiliation{$^3$LAPLACE (Laboratoire Plasma et Conversion
d'Energie), Universit\'e Paul Sabatier, 118 route de Narbonne, 31062
Toulouse Cedex 9, France}

\affiliation{$^4$Service de Physique de l'\'Etat Condens\'e, CNRS URA
2464, CEA -- Saclay, 91191 Gif-sur-Yvette, France}

\affiliation{$^5$Laboratoire de Physique Th\'eorique, Universit\'e Paul
Sabatier, 31062 Toulouse Cedex 4, France}

\affiliation{$^6$CNRS, Laboratoire de Physique Th\'eorique, F-31062
Toulouse, France}

\begin{abstract}
Fish schools are able to display a rich variety of collective states
and behavioural responses when they are confronted to threats.
However, a school's response to perturbations may be different
depending on the nature of its collective state. We use a previously
developed data-driven fish school model to investigate how the school
responds to perturbations depending on its different collective
states, we measure its susceptibility to such perturbations, and
exploit its relation with the intrinsic fluctuations in the school. In
particular, we study how a single or a small number of perturbing
individuals whose attraction and alignment parameters are different
from those of the main population affect the long-term behaviour of a
school. We find that the responsiveness of the school to the
perturbations is maximum near the transition region between milling
and schooling states where the school exhibits multistability and
regularly shifts between these two states. It is also in this region
that the susceptibility, and hence the fluctuations, of the
polarisation order parameter is maximal. We also find that a
significant school's response to a perturbation only happens below a
certain threshold of the noise to social interactions ratio.
\end{abstract}

\maketitle

\section{Introduction}

Fish schools behave as coherent entities and display complex emergent
properties such as coordinated motion, different ordered collective
states, and rapid escape manoeuvres when attacked by predators
\cite{Breder1951, Radakov1973, Partridge1982, Treherne1981,
  Domenici2004}. These group-level properties provide evolutionary
advantages to fish schools and arise through social interactions by
which individuals exchange information and perform specific
behavioural responses such as changing their direction and velocity,
or avoiding collision with group members or obstacles in the
environment \cite{Krause2002, Treherne1980, Foster1981, Magurran1990,
  Domenici1997, Herbert2011, Katz2011}. These interactions facilitate
the transfer of information between fish and their ability to
quickly respond to changes in the environment. For instance, when
some fish spot a predator they abruptly change their direction of
travel. Their close neighbours react in turn by changing their own
velocity so that the information gradually propagates through the
whole group, allowing all individuals to escape \cite{Krause2002}.
Experimental and theoretical works have shown that the same
interactions can lead a few individuals having salient information,
such as knowledge about the location of a food source or of a
migration route, to guide other group members and bias the resulting
direction of travel of the school \cite{Romey1996, Huse2002,
Couzin2005}. In all these situations, a small proportion of
individuals deeply influence the collective behaviour of the whole
group. This is a direct consequence of the asymmetry of behavioural
responses that exist between the perturbing or informed individuals
and the other group members. Indeed, the behaviour of a perturbing
or an informed fish is weakly influenced by the behaviour of its
neighbours. While the decisions of the neighbours are mostly
dictated by the behaviour of nearby fish.

Theoretical investigations have shown that not only the features of
local interactions among individuals but also the number and position
of neighbours to which a fish pays attention determine the patterns of
collective motion that emerge at the group level \cite{Aoki1982,
  Huth1992, Couzin2002, Viscido2005}. For instance, we have recently
shown in a data-driven fish school model that the relative weights of
attraction and alignment interactions between fish give rise to a
small number of specific collective states \cite{Gautrais2009,
  Gautrais2012, Calovi2014}: a {\it swarming} (disordered) state in
which fish aggregate without cohesion, with a low level of
polarisation of their velocity; a {\it schooling} (ordered) state in
which individuals are aligned with each other; a {\it milling}
(ordered) state in which individuals constantly rotate around an
empty core; and a {\it winding} (ordered) state, in which the group
exhibits an elongated phase characterised by a linear crawling
motion. However, according to whether fish pay equal attention to
their surrounding neighbours or focus their attention only of those
neighbours that are ahead of them \cite{Calovi2014} ({\it i.e.},
interactions depend on the neighbour angular position), the number
of collective states that can be reached by a school is different.
Only the swarming and schooling states survive when the behavioural
reactions of fish do not depend on the angular position of their
neighbours, while the full repertoire of collective states exists
when a front/back asymmetry in fish interaction is introduced.
Moreover the exploration of the model has shown that in the
transition region between milling and schooling states, the school
exhibits multistability and regularly shifts from schooling to
milling for the same combination of individual parameters, a
property that was also reported in the model of Couzin {\it et al}.
\cite{Couzin2002} and in experimental observations on groups of
golden shiners \cite{Tunstrom2013}.

While the past literature on schooling models has been devoted to a
better understanding of the link between interaction rules and
collective behaviours, less attention was paid to the group response
and to its sensitivity to external perturbations \cite{Aoki1982,
  Couzin2002, Viscido2005, Hemelrijk2008, Gautrais2009, Hemelrijk2010,
  Calovi2014, Kolpas2013, Aureli2010}. In particular, one may wonder
if a small number of perturbed or informed individuals could trigger
the same response whatever the collective state of the school.

In the present work, we extensively study a previously developed
data-driven model \cite{Gautrais2012}, which has been validated
previously on actual experiments on {\it Khulia mugil} and was able
to describe quantitatively several properties: individual
trajectories of a single fish in a tank (validating in particular
the noise and friction terms; see the Model section below),
interaction between 2 to 30 fish, diffusion properties, mean
distance between fish, mean fish alignment/polarisation vs the
velocity or the number of fish in the tank. In another work
\cite{Calovi2014}, the phase diagram of the model \emph{without the
tank boundaries} (in free space) was studied as a function of the
attraction and alignment parameters, reproducing several collective
states observed in actual fish schools (see below). The excellent
qualitative and quantitative accuracy of the model in describing
real fish schools in a tank is a good motivation to consider it as a
fair description of fish moving in a free space (in particular their
response to a perturbation) for which relevant experiments would be
much harder to implement.

The present work hence addresses the \emph{response properties} of a
fish school in free space, using the model as presented in
\cite{Calovi2014}, and that we will briefly review below for
completeness, but also to emphasise the crucial role played by the
fish anisotropic angular perception of their environment. We
investigate how a single or a small number of perturbed individuals
affect the long-term behaviour of a school. In particular, we study
how the school responds to perturbations depending on its different
collective states, and introduce their \emph{susceptibility} to such
perturbations. We relate the fish school response quantified by
these susceptibilities to the fluctuations (for instance, of the
polarisation order parameter) already existing in the unperturbed
fish school. This deep connection between response to a perturbation
and intrinsic fluctuations without perturbation, although well
understood and studied in the context of physical systems, is less
familiar in the present context of fish schools, but equally
applies. It is thus also one of the main purpose of the present work
to illustrate and study qualitatively and quantitatively this
general connection. In addition, we also exploit the fact that
fluctuations and hence adequate susceptibilities are maximum (and
would diverge in an infinite systems) at the transition between two
(collective) states separated by a continuous phase transition.

We then first analyse the school susceptibility in the absence of
any perturbation and its relation to fluctuations and to the
identification of transition lines between the different collective
states. We also explore the influence on the group behaviour of a
perturbing fish with an independent set of attraction and alignment
parameters, while keeping the main population in the high
susceptibility region, in order to determine the perturbations which
have the higher impact. We finally explore the model parameter space
to determine how the relative weighting of attraction and alignment
of fish affects the school responsiveness to perturbations. Finally,
we discuss the implications of our work for real fish schools.

\section{Model}

\begin{figure}[!htb]
\begin{center}
 \includegraphics[width=0.9\linewidth]{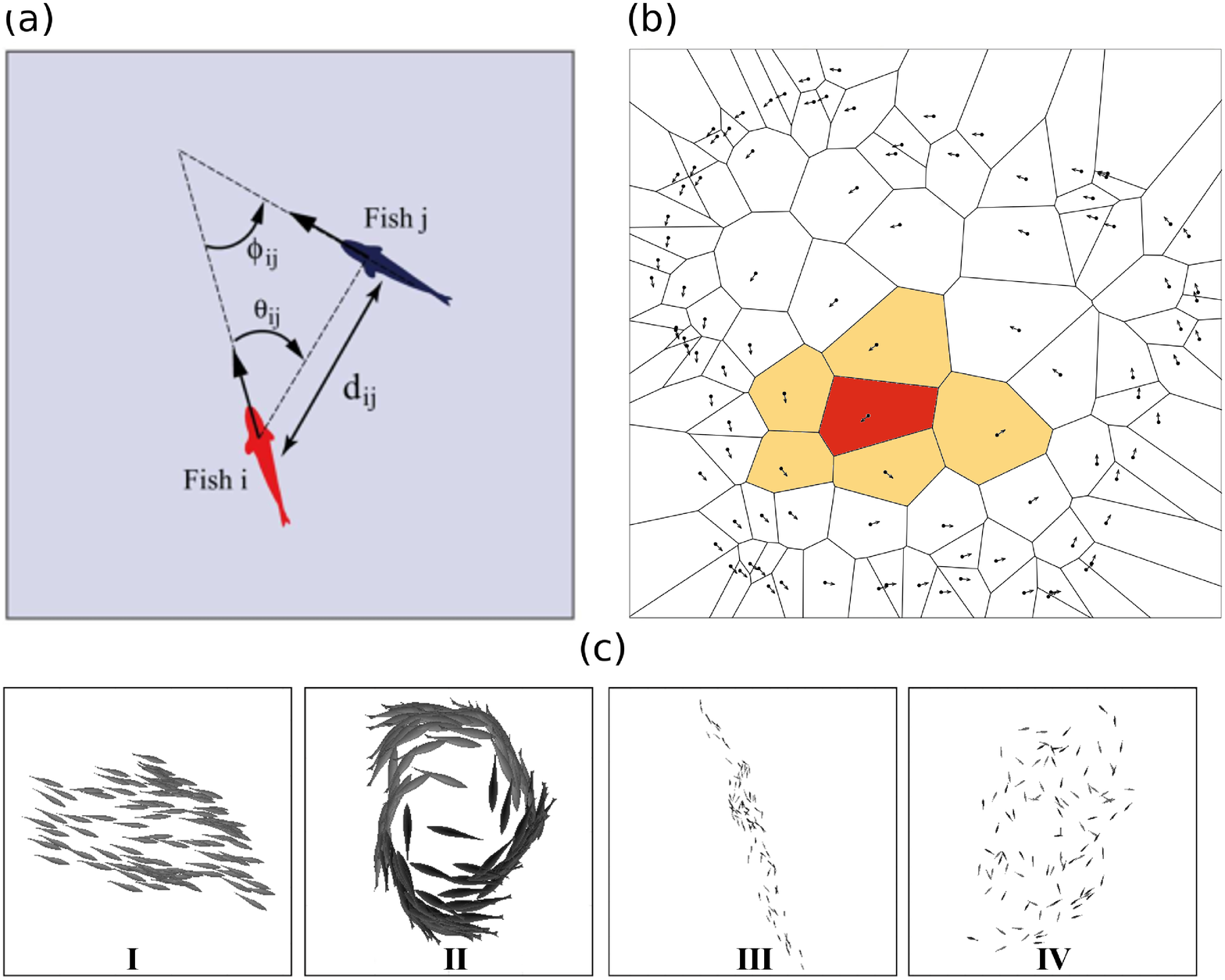}
 \caption{Graphical representation of fish interactions used in the
   model. (a) $d_{ij}$ is the distance of fish $j$ from fish $i$;
   $\phi_{ij}$ is the relative orientation of fish $j$ compared to
   fish $i$; $\theta_{ij}$ is the angle between the angular position
   of fish $j$ with respect to fish $i$. (b) Illustration of the
   Voronoi neighbourhood; arrows indicate fish headings. A focal fish
   is displayed in red and his Voronoi neighbours in orange. (c)
   Snapshots of typical configurations for the 4 distinct states
   displayed by the model.}
\label{Geometrical}
\end{center}
\end{figure}

The present model was originally proposed by Gautrais {\it et al}.
\cite{Gautrais2009, Gautrais2012} to describe the coordination of
movements in groups of {\it Khulia mugil} through the use of
stochastic equations of motion
for their angular velocity $\omega_i=\ud\phi_i/\ud t$, while fish
move with a constant speed along their angular direction given by
$\phi_i$. In a previous work \cite{Calovi2014}, we have introduced a
non-dimensionalised version of the model in which we included an
angular modulation of the strength of interactions between a fish
and its neighbours according to their angular position, a property
also supported by experiments \cite{Gautrais2009, Gautrais2012}, but
of negligible consequence for fish in a not too large tank. This
angular modulation breaks the symmetry of interactions between fish
in front and those in the back. For completeness, we briefly recall
the main ingredients of the model, where the individual
 angular velocity $\omega_i$ evolves according to the
following non-dimensional stochastic differential equation:
\begin{equation}
\alpha \ud \omega_i(t) = -  \left[\omega_i(t) -  \omega_i^{*}(t)\right]\ud t + \ud W_i(t),
\label{AngularSpeed}
\end{equation}
where $\alpha$ can be understood as an angular inertia term,
$\omega_i^{*}$ is the response function resulting from the
interaction with the neighbouring fish (see figure
\ref{Geometrical}(a)) and $\ud W_i(t)$ refers to a random variable,
uncorrelated in time, and uniformly distributed in the interval
$[-1,1]$ (times $\sqrt{\ud t}$). As a consequence of the large
number theorem, such a uniformly distributed noise has exactly the
same effect as the usual Gaussian noise (Wiener process) in the long
run and in the limit of a time step $\ud t\to 0$, with the benefit
of being much faster to implement numerically.

The interaction is described by the normalised linear superposition
of pair interactions between the focal fish and the first shell of
Voronoi neighbours (see figure \ref{Geometrical}(b)) as follows:
\begin{equation}
\omega_i^{*} =\frac{1}{N_i}\sum_{j \in
  V_i}\left[1 +\cos( \theta_{ij} )\right]
  \left[\beta \sin \phi_{ij} + \gamma d_{ij} \sin
  \theta_{ij}\right],
\label{Omega*}
\end{equation}
where $\theta_{ij}$ is the angular position between the focal fish
$i$ and the neighbour $j$, $N_i$ refers to the number of
neighbouring fish in the first shell of the Voronoi tessellation
$V_i$. The following terms correspond respectively to the alignment
and attraction ``forces'', with $\beta$ and $\gamma$ controlling
their intensities. $\phi_{ij}=\phi_j-\phi_i$ is the heading angle
difference and $d_{ij}$ is the non-dimensional distance between $j$
and the focal fish $i$. Originally, when investigating fish
interactions, Gautrais {\it et al.} \cite{Gautrais2012} considered
different kind of neighbourhoods to combine fish interactions. It
was shown that besides the Voronoi neighbourhood, other choices were
compatible with the experimental data in a tank, in particular the
interactions between the $k$ ($k\approx 6-8$) nearest neighbours,
which is remarkably similar to the average number of neighbours
contained in the first shell of the Voronoi tessellation ($\langle
N_i\rangle=6$, exactly, in two dimensions).

In equation \ref{Omega*}, the prefactor $A_{ij}=1
+\cos(\theta_{ij})$ modulates the amplitude of the interaction
between fish $i$ and $j$, as a function of the angle of view of the
former. It is larger if $j$ is ahead $i$ than behind, and is hence
maximum for $\theta_{ij}=0$ ($A_{ij}=2$), minimum for
$\theta_{ij}=\pi$ ($A_{ij}=0$), while its angular average is
$\langle A_{ij}\rangle=1$. This term also introduces a strong
asymmetry between the force exerted by $j$ on $i$ and the one
exerted by $i$ on $j$, and hence breaks the (Newtonian)
action-reaction principle which is most familiar in the context of
purely physical force, like gravitation. Apart from the technical
fact that the alignment and attraction forces become non
conservative (\emph{i.e.} not deriving from a potential energy), it
has the practical and important consequence of allowing for the
milling and winding phases, both observed in actual fish schools in
the ocean (quasi free space).

In \cite{Calovi2014}, we have shown that varying the parameters
$\beta$ and $\gamma$ of the model strongly affects the school
behaviour, leading to four distinct collective states (see the
complete phase diagram in \cite{Calovi2014}): (I) schooling, (II)
milling, (III) elongated winding state and (IV) swarming, all of
which can be visualised in figure \ref{Geometrical}(c).


\section{Quantification of collective behaviours and responses to perturbations}

We now proceed to describe the tools used to characterise the
collective states and to measure the susceptibility and the school
response to perturbations. We also describe the numerical details
used for the simulations herein.

\subsection{Order parameters}
The aforementioned states can be quantified by two order parameters:
(1) the polarisation order parameter which provides a measure of how
aligned the individuals in a group are:
\begin{equation}
P= \frac{1}{N} \left|\sum_{i=1}^{N} \frac{\vec{v}_i}{v}\right|.
\label{P}
\end{equation}
$P$ takes values between 0 (no alignment on average) and 1 (all fish
are aligned); and (2) the rotational or milling order parameter
which provides a measure of the milling behaviour. It is the
absolute value of the normalised angular momentum:
\begin{equation}
M= \frac{1}{N} \left| \sum_{i=1}^{N}{\frac{ \vec{r}_i {\times}
    \vec{v}_i }{|\vec{r}_i| v}}\right|,
\end{equation}
where $|\vec{v}_i|=v=1$ in the non-dimensional version of the model.
$M$ takes values between 0 (no collective rotation) and 1 (strong
collective rotation of the school). The analysis on the transition
line between the schooling and milling phases \cite{Calovi2014}
shows that it obeys a simple functional form $\beta=A
\sqrt{\gamma}+B$, and that it is independent of the angular inertia
term $\alpha$ from equation (\ref{AngularSpeed}) considering values
of equivalent of constant speeds between $0.4$ and $1.2$\,m/s.

\subsection{Susceptibilities and fluctuations}

\label{definition}

In many physical systems, some physical quantity may be coupled
linearly and influenced by an external ``field''. For instance, in a
magnetic system, the atomic or electronic spins are coupled and tend
to align along the direction of an external magnetic field.
Similarly, an elastic medium (a spring or a rubber) can be elongated
or compressed by exerting an external force. In general, the linear
response of such a quantity $P$ (for instance, the spin/fish
polarisation -- \textit{i.e.} the average spin/fish direction) to a
small change in the associated external field $h$ is quantified by
introducing the $P$-susceptibility
\begin{equation}
P(h)=_{h\to 0}\chi h+...,\qquad \chi =\frac{\partial P}{\partial
h}_{|h=0}.
\label{eqlin}
\end{equation}

In the present context of the dynamics of fish school, the change in
the order parameter $P$ (or $M$) to a small perturbation (of a
nature detailed in the two next sections) will give a first
characterisation of the response of the system which will be
presented in section \ref{quantpert} and \ref{grprespsec}.

Moreover, for a system at equilibrium or in a stationary state
associated to an energy functional (an Hamiltonian), the
fluctuation-dissipation theorem (FDT) \cite{Marconi2008} states that
there exists a direct relation, in fact an \emph{exact equality} up
to a constant factor, between the $P$-susceptibility, as defined
above by means of a small perturbing field, and the (thermal)
\emph{fluctuations} of $P$ at equilibrium
\begin{equation}
  \chi = N \left[\langle P ^2 \rangle -
\langle P \rangle^{2}\right]=\frac{\partial P}{\partial h}_{|h=0},
\label{eqsusc}
\end{equation}
where $N$ is the number of particles, and the angular brackets refer
to the average over time of the corresponding quantity, {\it e.g.}
$\langle P
\rangle=\lim_{t_0\to\infty}\frac{1}{t_0}\sum_{t=0}^{t_0}P(t)$. In
practice, in numerical simulations, $t_0$ is of course finite, and
is taken as large as computation time permits. In addition, the
susceptibility is also averaged over as many different
samples/initial conditions as computationally possible. Note that
the neighbouring spins/fish of a given spin/fish exert an effective
magnetic/alignment field on the latter, making the connection
between response and fluctuations very natural. In addition, the
noise in equation (\ref{AngularSpeed}) formally plays exactly the
same role as the thermal noise in physics.

This powerful FDT has several very important implications, apart
from the clear physical insight gained on the relations between the
fluctuations and the response of a system. For instance, in
numerical simulations (molecular dynamics or Monte Carlo), it is
much easier and much more precise to measure the susceptibility from
the fluctuations of the order parameter $P$ (see the first equality
in equation (\ref{eqsusc})), rather than applying a small field $h$,
waiting for equilibrium to settle, measuring the (small)
perturbation on $P$, and ultimately trying to extrapolate to $h=0$
(see the second quality in equation (\ref{eqsusc}), and the original
definition of equation (\ref{eqlin})). In the present study of fish
schools, we will first measure the susceptibility via the
fluctuations of the system (first equality in equation
(\ref{eqsusc})), and will in particular compare this susceptibility
to the change of the order parameter under the addition of a few
perturbing fish to the school (effectively acting as a small
perturbing field).

As already mentioned, the asymmetric forces resulting from the
$A_{ij}=1 +\cos( \theta_{ij})$  term breaks the Newtonian
action-reaction principle and forbids the existence of an underlying
Hamiltonian. However, the FDT has been generalised in out of
equilibrium situations, including in cases where the system,
although in a stationary state, is not formally described by an
energy functional (see \cite{Marconi2008} for a review). Hence, the
connection between the fluctuations of the school polarisation and
the polarisation response to a perturbation made in the present work
appears very natural, and will lead to important biological
implications.

In the same manner as we just defined the polarisation
susceptibility, we can define the milling susceptibility $\chi_m$
associated to the milling order parameter
\begin{equation}
  \chi_m = N \left[ \langle M ^2 \rangle - \langle M \rangle^{2}\right].
\label{eqsuscm}
\end{equation}
Again, it should be intimately related to the change of $M$ under a
small perturbation defined hereafter, a relation which will be
illustrated in the Results section and in the supplementary figures.

Finally, in the physical context, the susceptibility and hence
fluctuations are known to \emph{diverge} (at least in the limit of
an infinite system $N\to\infty$) exactly at the critical point
between two phases separated by a continuous (second order) phase
transition involving the considered order parameter. In the present
context of fish schools, we will be naturally interested in the
behaviour of the susceptibility and fluctuations near transition
lines, in particular near the schooling-milling transition. The
maximum of the susceptibility (characterised by fluctuations
\emph{or} response) as a function of the model parameters hence
provides an alternative identification of the transition lines which
will be illustrated extensively in section \ref{nopertsusc} and in
the supplementary figures.

\subsection{Quantifying school response to perturbations}

To detect behavioural changes, we compare the average values of the
polarisation and milling order parameters for a given set of alignment
and attraction parameters of the unperturbed case, with the new
average value given by the simulations with one or more perturbing
fish. These results are presented for different combinations of the
attraction and alignment parameters: complete parameter space scans,
cross-sections where the attraction parameter is kept fixed, and
lastly, by the set of parameters which describe the transition between
the milling and schooling state.

\subsection{Simulations}

We investigate the long-term consequences on the resulting school
behaviour of a small number of perturbing fish that differ from the
main population by having a different combination of attraction and
alignment parameters. Henceforth, we call $N_p$ the number of
perturbing fish and $N_m=N-N_p$ the main population of a $N$ fish
school. Accordingly ($\gamma_p,\beta_p$), and ($\gamma_m,\beta_m$),
are the attraction and alignment parameters respectively of the
perturbing fish and the main population. All simulations were run in
an unbounded space with 400 random initial conditions for 1000
non-dimensional time units, where the first half was discarded to
remove transient states. A simple Euler's method integration with a
time step of d$t=1.44{\times} 10^{-3}$ proved sufficient to avoid
numerical imprecisions. Simulations results shown in the next sections
were performed with $N=100$ or 200, meanwhile the number of perturbing
fish $N_p$ may vary from 1 to 9 depending to the studied conditions.

\section{Results}

\subsection{Susceptibility of a group of fish without perturbations}
\label{nopertsusc}

\begin{figure}[!Htb]
\begin{center}
 \includegraphics[width=0.65\linewidth]{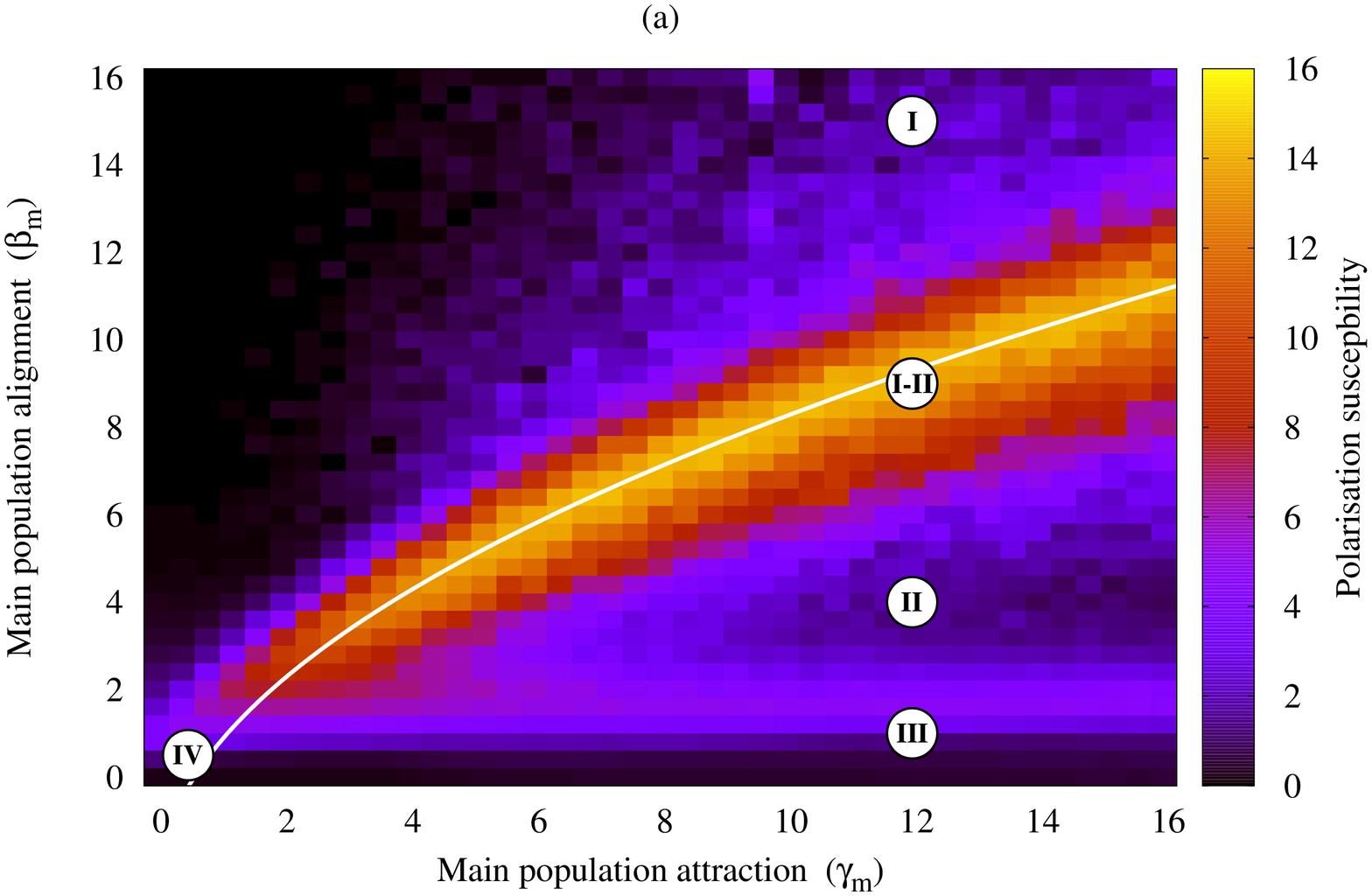}
 \includegraphics[width=0.65\linewidth]{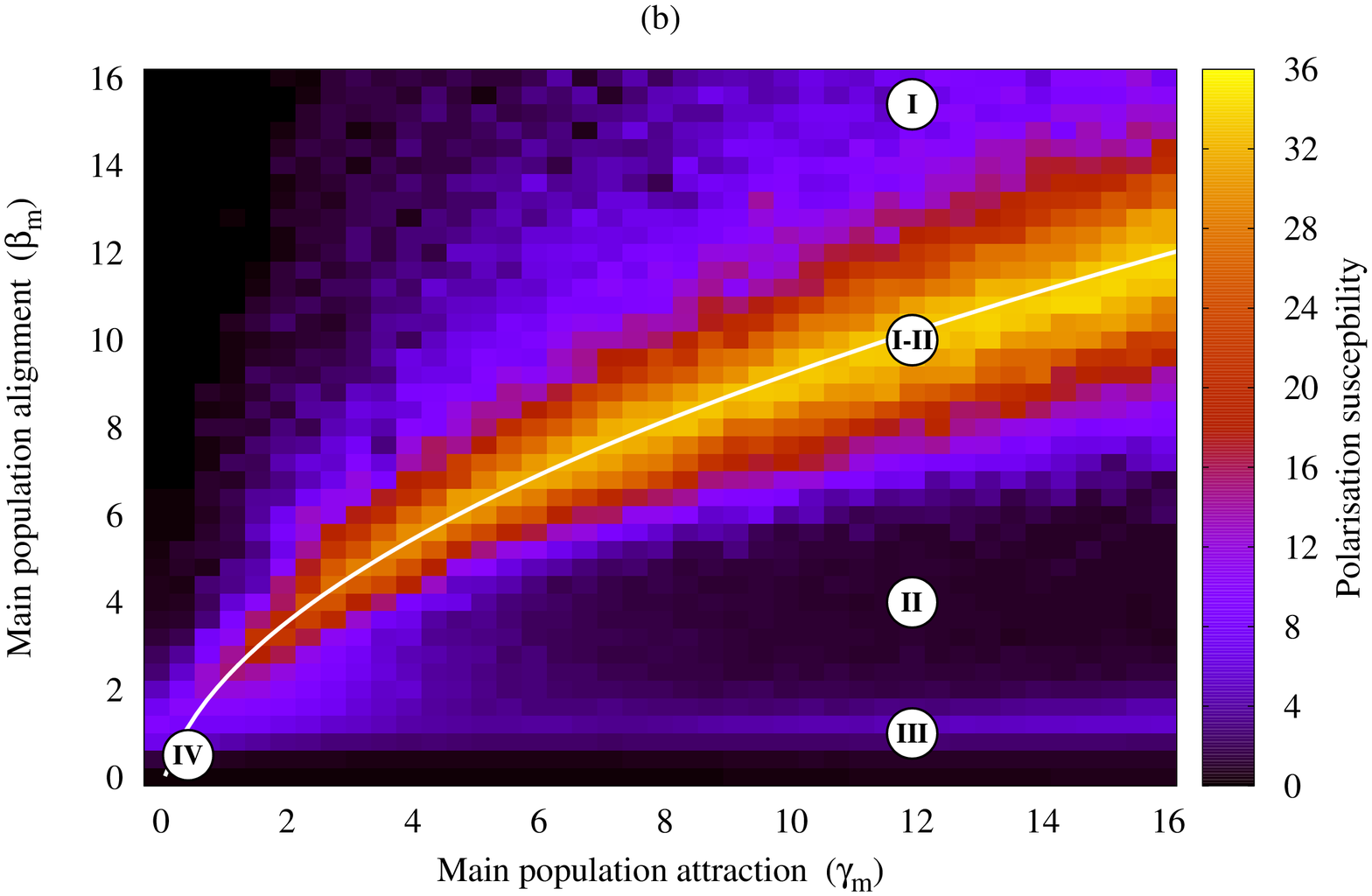}
 \caption{Susceptibility of unperturbed fish school simulations for
   100 (a) and 200 (b) fish.  The susceptibility is calculated according
   to equation (\ref{eqsusc}) for different values of the attraction
   and alignment parameters. Each data point represents an average over
   400 simulations with random initial conditions. The white lines
   following the peak of susceptibility represents the function that
   fits the schooling/milling transition line as reported in
   \cite{Calovi2014}. The circled numbers indicate the 4 different
   collective states (I) schooling, (II) milling, (III) winding,(IV)
   swarming and the transition zone between schooling and milling
   (I-II).}
\label{Suscep}
\end{center}
\end{figure}

Having defined the susceptibility previously via the order parameter
fluctuations (see equation (\ref{eqsusc})), we have calculated the
polarisation susceptibility values for different combinations of
attraction and alignment parameters in a group of $N=N_{m}=100$ and
$N=N_{m}=200$ fish respectively, shown in figures \ref{Suscep}(a)
and \ref{Suscep}(b). One can clearly see that the highest values of
susceptibility correspond perfectly to the fitted transition line
between the schooling and milling regions. This indicates that the
transition region is a good candidate to test different types of
perturbing fish. Together with the susceptibility, other statistics
of the unperturbed simulations were also computed, such as
polarisation and the milling order parameters, so that we can use
some regions of this map as a baseline to measure the impact of
perturbing fish on the school behaviour.

In section \ref{grprespsec} (see in particular figure \ref{P100}),
and in the supplementary material (by considering the milling
susceptibility $\chi_m$), we will indeed find a very strong
correlation between the susceptibility computed from the
fluctuations in the unperturbed system, and the response properties
of the system under small perturbations.

\subsection{Quantification of perturbations}
\label{quantpert}

\begin{figure}[floatfix]
\begin{center}
 \includegraphics[width=0.9\linewidth]{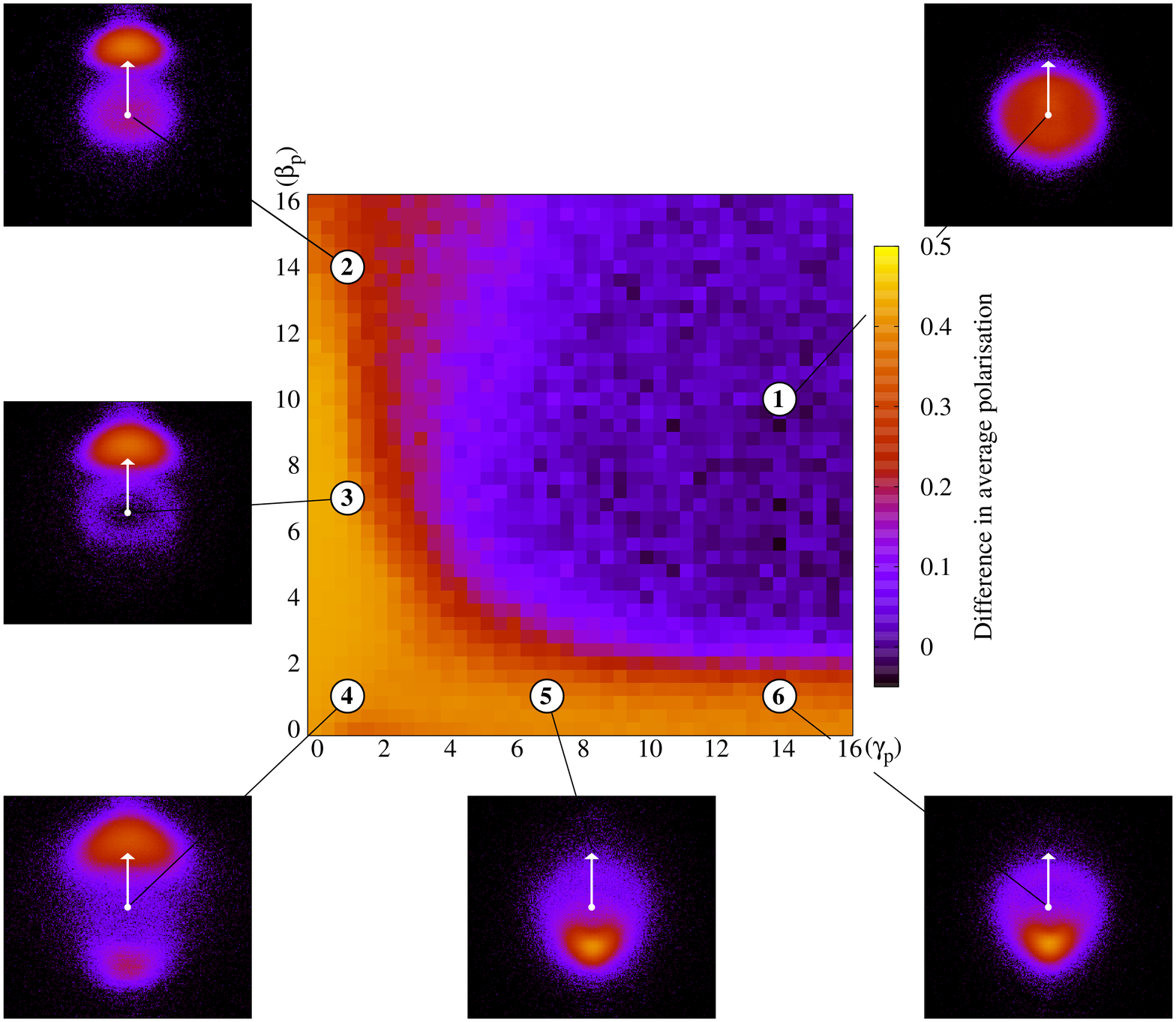}
 \caption{Each data point represents the difference in the average
   polarisation caused by a single perturbing fish, while varying its
   attraction ($\gamma_p$) and alignment ($\beta_p$) parameters, in
   comparison to the unperturbed case located on the transition region
   ($\gamma_m= 14$, $\beta_m= 10$). The six insets shown here
   represent the location distribution of the perturbing fish
   positions in relation to the group's centroid (white circle), and
   reoriented in according to the average direction movement of the
   group (white arrow).}
\label{PertScan}
\end{center}
\end{figure}

We now proceed to investigate the impact of various combinations of
attraction and alignment parameters of a single perturbing fish on
the resulting group behaviour. The main population is located in the
transition region ($\gamma_m=14$, $\beta_m=10$) while the parameters
$\gamma_p$ and $\beta_p$ of the perturbing fish both vary in the
interval $[0,16]$ with a $0.4$ step, forming an uniform $41{\times}41$ grid
in the parameters space. Figure \ref{PertScan} shows the resulting
difference in the average polarisation $P$. Note that since the main
population is located in the transition region, the values of the
polarisation $P$ and milling $M$ parameters fluctuate around $0.5$.
This means that in the case where the group changes to an almost
perfect schooling state ($P\approx 1$), the maximum difference is
$0.5$. Figure \ref{PertScan} shows that this change to a schooling
state happens for both low attraction and low alignment values
($\gamma_p$ and $\beta_p < 2$).

One can easily understand that for low values of $\gamma_p$,
corresponding to a weak attraction, the perturbing fish will not
remain as close to the group as the others, forcing them to follow
it, and in doing so, increasing the polarisation of the group. For
low values of $\beta_p$, fish can have the same attraction
parameters as the main population, but the weak alignment disrupts
the milling intermittence completely. The six insets in figure
\ref{PertScan} represent the distribution of perturbing fish
locations with respect to the group's centroid (white circle), and
reoriented according to the average direction movement of the school
(white arrow). Inset 1 shows the unperturbed case
($\gamma_m=\gamma_p$ and $\beta_m=\beta_p$), where we can see that
the perturbing fish has an equal distribution all around the school.
One can see in insets 2 to 4 (and in supplementary videos
\ref{inset3} and \ref{inset4}) that when there is a low attraction
($\gamma_p=1$) the perturbing fish stays most of the time ahead of
the group. This ``leading'' behaviour in which the school is
attracted by the perturbing fish, but not the opposite, is a
consequence of the smaller attraction of the perturbing fish
compared to the main population ($\gamma_{p} < \gamma_{m}$). Higher
values of $\gamma_p$ combined with a weak alignment ($\gamma_p=7$ or
14, and $\beta_p=1$) lead the perturbing fish to stay usually behind
the group's centroid (insets 5 and 6 and supplementary video
\ref{inset6}). It is also important to highlight that in the latter
case the perturbing fish is much closer to the group's centroid in
comparison to the conditions shown in the insets 2 through 4. The
reason for the perturbing fish to remain close and behind to the
school's centroid is quite simple. The high attraction insures that
the perturbing fish remains close to the school, but with a low
alignment, it is unable to cope with the directional changes of the
other fish. As a consequence, it remains behind the school.

We also performed the same systematic analysis of the impact of a
perturbing fish on the group behaviour when the main population is
in the schooling or the milling state. Supplementary figures
\ref{Pert_School} and \ref{Pert_Mill} show the results of these
simulations. When the main population is located in the schooling
region ($\gamma_m=4$, $\beta_m=14$) a perturbing fish causes almost
no change to the group behaviour. When the population is located in
the milling region ($\gamma_m=14$, $\beta_m=4$) one can observe a
change from milling to schooling for low parameter values of the
perturbing fish ($\gamma_p$ and $\beta_p < 2$) and intermittent
transition between schooling and milling appear for low attraction
and high alignment values ($\gamma_p<2$ and $\beta_p > 10$). As
explained previously, these effects in the milling region are only
due to a fish which is not able to remain close to the group,
forcing its neighbours to follow him, and in doing so, disrupting
the mill, as shown in figure \ref{PertScan} (insets 2 to 4). It is
worth noting that: (1) the only observed change in the group
behaviour resulting from the presence of a perturbing fish is a
transition to schooling; (2) only a perturbing fish with very low
attraction values ($\gamma_{p}\approx 0$) is able to disrupt a group
engaged in a milling state; any other behaviour of the perturbing
fish has no effective impact on the group tendency to rotate.

\subsection{Group response to perturbations}
\label{grprespsec}

\begin{figure}[!htb]
\begin{center}
  \includegraphics[width=0.95\linewidth]{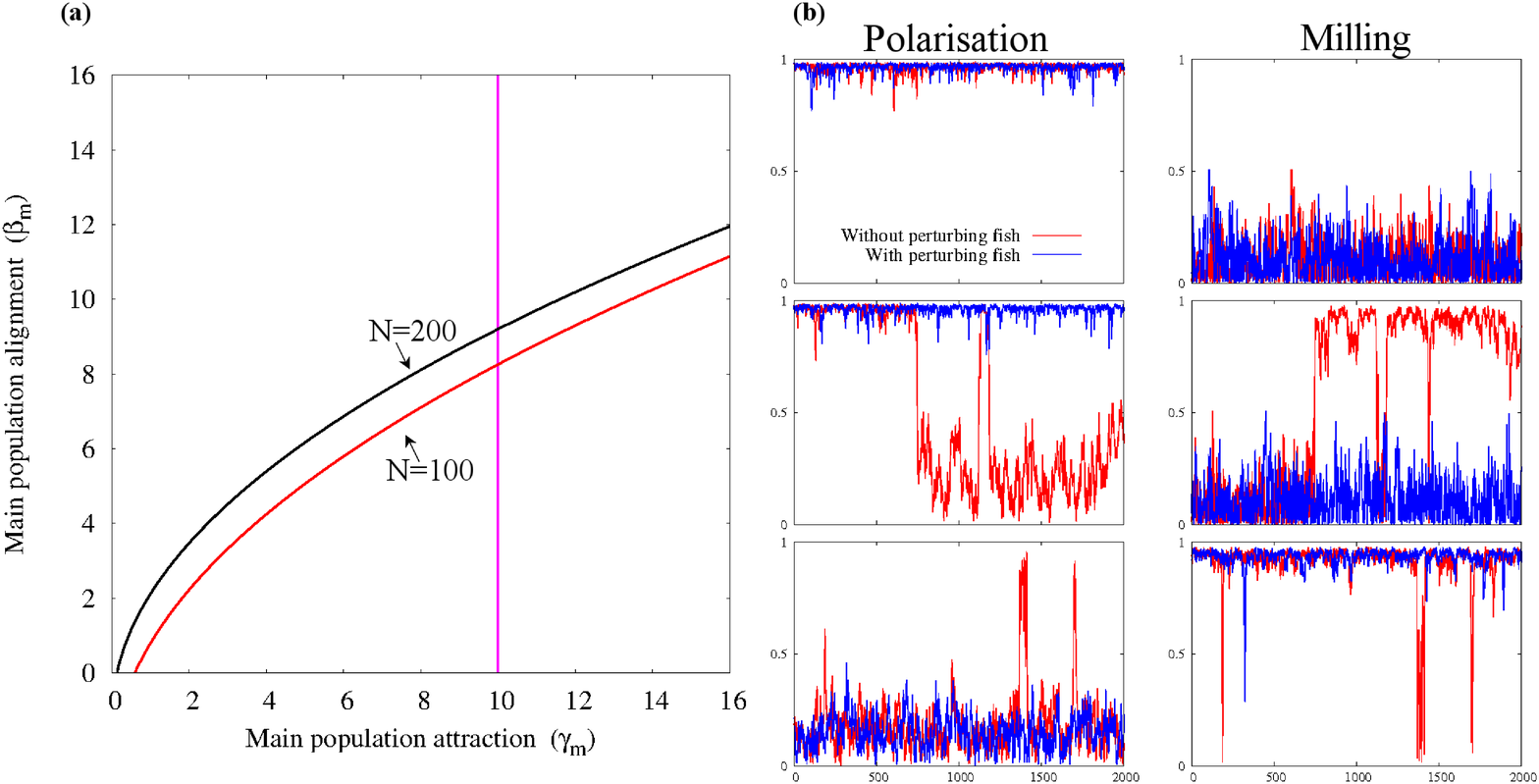}
\caption{(a) Transition functions between schooling and milling fitted
  for simulations with 100 and 200 fish $\beta_m=A \sqrt{\gamma_m}+B$
  where [$A=3.22$, $B=-2.23$] for 100 fish and [$A=3.28$, $B=-1.17$]
  for 200 fish. (b) Time series of polarisation $P$ and $M$ milling
  parameters when the main population is in 3 different states:
  schooling ($\gamma_m=4$, $\beta_m=14$), milling ($\gamma_m=14$,
  $\beta_m=4$) and the transition between both states ($\gamma_m=14$,$
  \beta_m=10$), for unperturbed condition (red line) and with a single
  perturbing fish introduced in the group (blue line, $\gamma_p=10$,
  $\beta_p=1$).}
\label{impact}
\end{center}
\end{figure}

Considering the results shown in figure \ref{PertScan}, we have chosen
a configuration of parameters for the perturbing fish ($\gamma_p=14$
and $\beta_p=1$) which lies in the winding region (inset 6). The
choice is motivated to prevent the perturbing effects to be the simple
consequence of non gregarious fish ($\gamma_{p}\approx 0$), like the
situations depicted in insets 2 to 4 in figure \ref{PertScan}. We now
proceed to analyse the group response to this perturbation for
different configurations of the main population parameter space. We
focus on a cross section of the parameter space keeping a fixed value
of $\gamma_m=10$ and varying $\beta_m$ in the interval $[0,16]$
represented by the vertical purple line in figure
\ref{impact}(a). Highlighted in figure \ref{impact}(b) are typical
time series of the polarisation and milling order parameters in the
schooling region (I), the transition region (I-II) and the milling
region (II). One can see that in the transition region, simulations
with the perturbing fish display a change to a purely schooling
behaviour.

We can now investigate how an increasing number of perturbing fish
affect group behaviour. We performed simulations with $N_p=1$, 3 ,
5, 7, and 9 perturbing fish (and $N_m=99$, 97, 95, 93, and 91, when
$N=100$ fish, and $N_m=199$, 197, 195, 193, and 191, in simulations
with groups of $N=200$ fish).

\begin{figure}[!htb]
\begin{center}
 \includegraphics[width=0.7\linewidth]{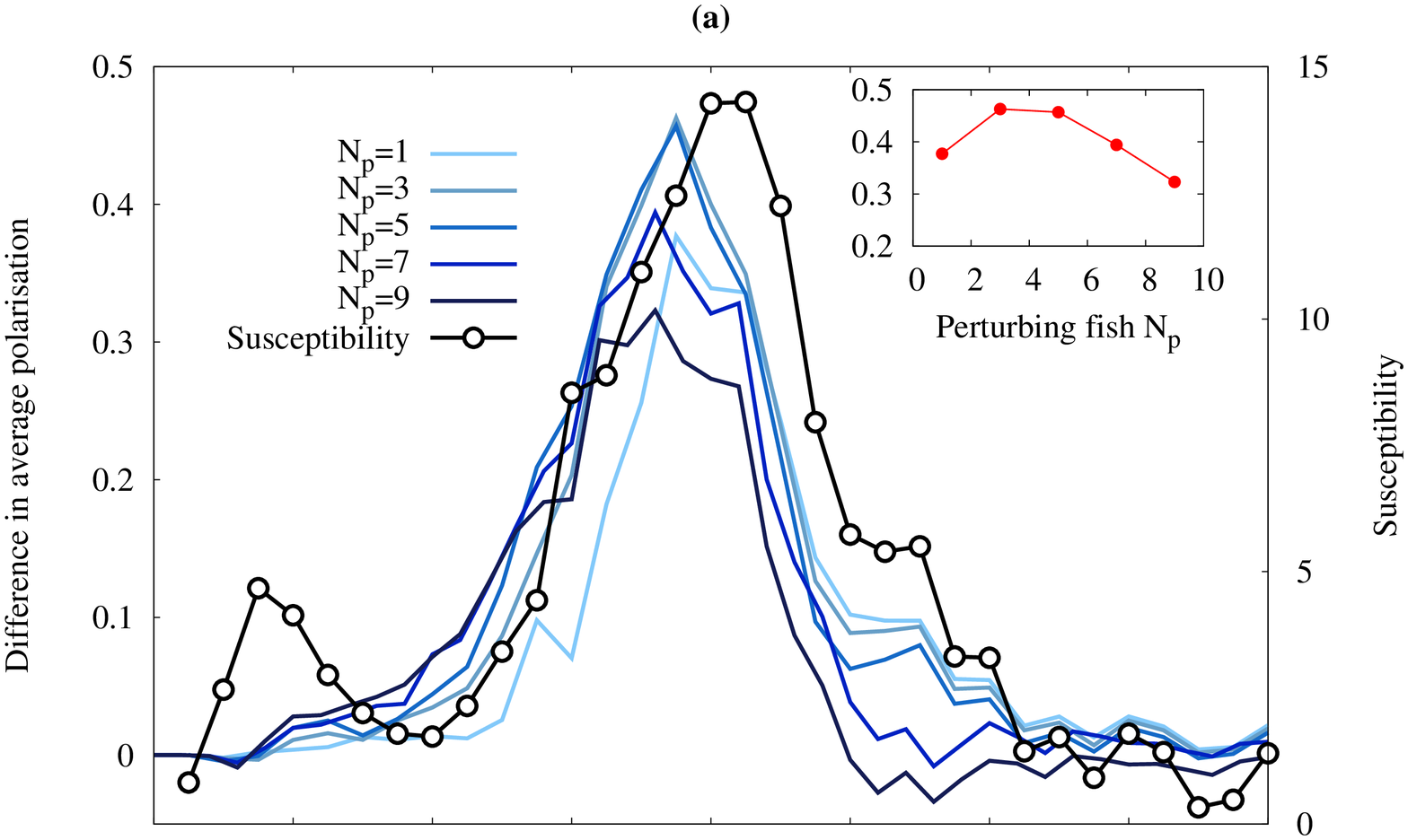}\\
 \includegraphics[width=0.7\linewidth]{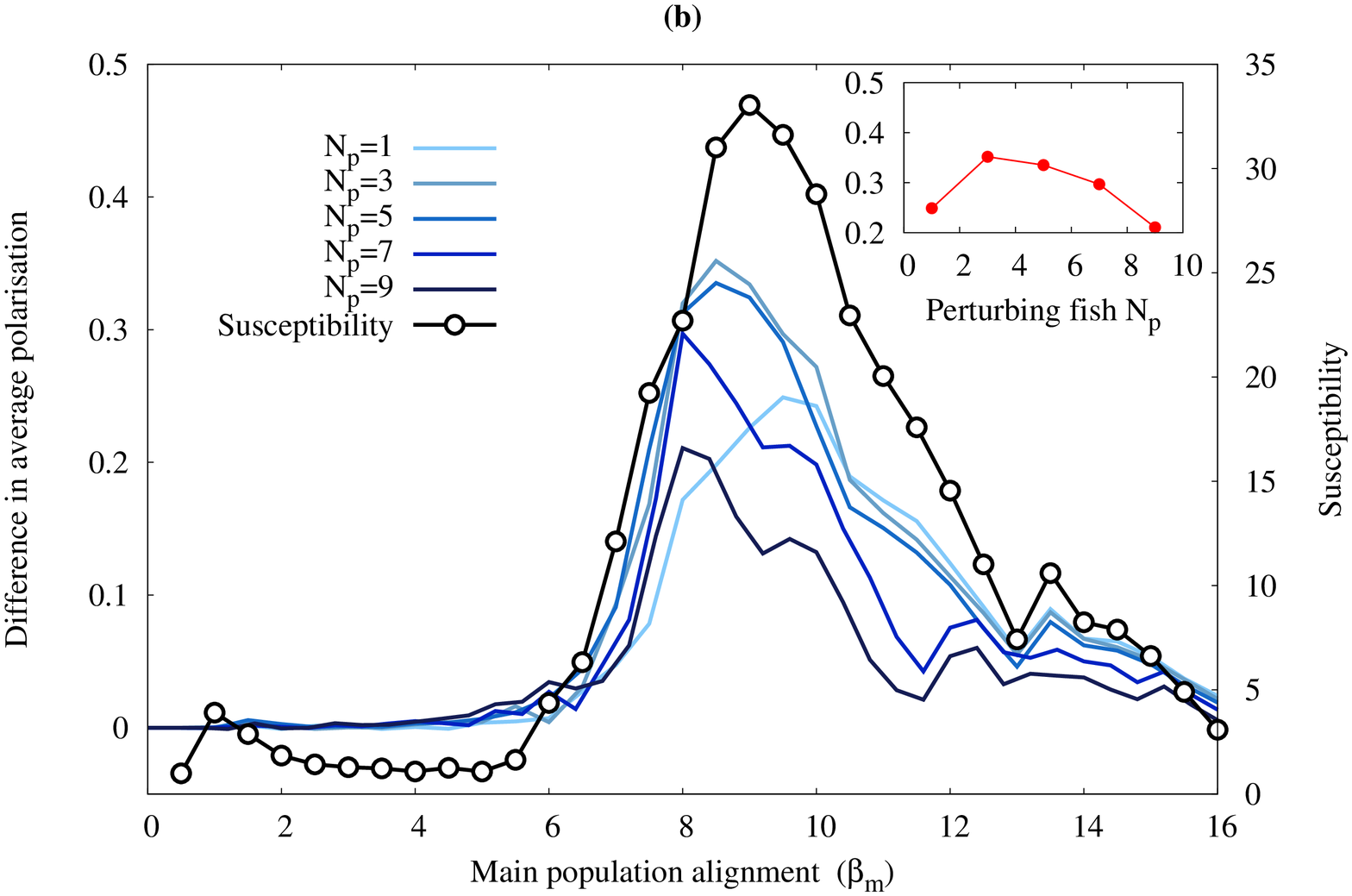}
 \caption{Difference in the average polarisation (left axis) in groups
   of 100 (a) and 200 (b) fish, with 1, 3, 5, 7, and 9 (light to dark
   blue) perturbing fish ($\gamma_p= 14$, $\beta_p= 1$), as a function
   of the alignment parameter of the main population ($\beta_m
   [0,16]$), keeping the attraction parameter of the main population
   to a constant value ($\gamma_m=10$). The black line represents the
   susceptibility values (right axis) for the unperturbed
   condition. The insets show the maximum difference in average polarisation
   as a function of the number of perturbing fish.}
\label{P100}
\end{center}
\end{figure}

Figure \ref{P100} shows the resulting difference in the average
polarisation induced by the perturbation in comparison to the
unperturbed condition in groups of 100 and 200 fish respectively. In
both cases the resulting change in the group polarisation and the
susceptibility follow a similar pattern reaching a peak in the
transition region. A smaller peak can also be seen for values of
  low $\beta_m$. This peak is related to the transition from the
  milling zone to the winding region.

Increasing the number of perturbing fish leads to an increase of the
group polarisation (linear perturbation regime) up to a saturation
value ($3<N_p< 5 $). Ultimately, it shows a steady decline with more
perturbing fish. This happens due to the fact that the perturbing
fish have a lower alignment parameter value, meaning that after the
initial perturbing effect, they will have a negative impact (non
linear perturbation regime) in the average polarisation of the
school. While 100 and 200 fish simulations show a very similar
pattern of change in group polarisation for all quantities of
perturbing fish, simulations with 200 fish display a systematic
lower response due to the perturbing fish.

One might notice the difference in two orders of magnitude between
difference in average polarisation and susceptibility. This comes from
the fact that the susceptibility is proportional to $N P^2$, while
the difference in average polarisation is only proportional to
$P$. Also, our main concern here is to use the susceptibility as a
reference point to which the group responses are compared.

Having established that the transition zone is the region of the
parameter space in which a fish group displays the highest
responsiveness to perturbations, we have studied the group response
throughout this region. As previously seen \cite{Calovi2014}, the
transition between schooling and milling follows the functional form
$\beta_m=A \sqrt{\gamma_m}+B$, where $A$ and $B$ fitted the
parameter space data in which the school presented both polarisation
and milling parameter values above 0.8 more than 40\% of the time
(see figure \ref{impact}). We can systematically vary the attraction
parameter $\gamma_m$ within the range $[0,16]$ and determine the
parameter $\beta_m$ estimated by this procedure. In the following
analysis, the parameters of the single perturbing fish are kept
unchanged ($\gamma_p=14$ and $\beta_p=1$).

\begin{figure}[!htb]
\begin{center}
 \includegraphics[width=0.9\linewidth]{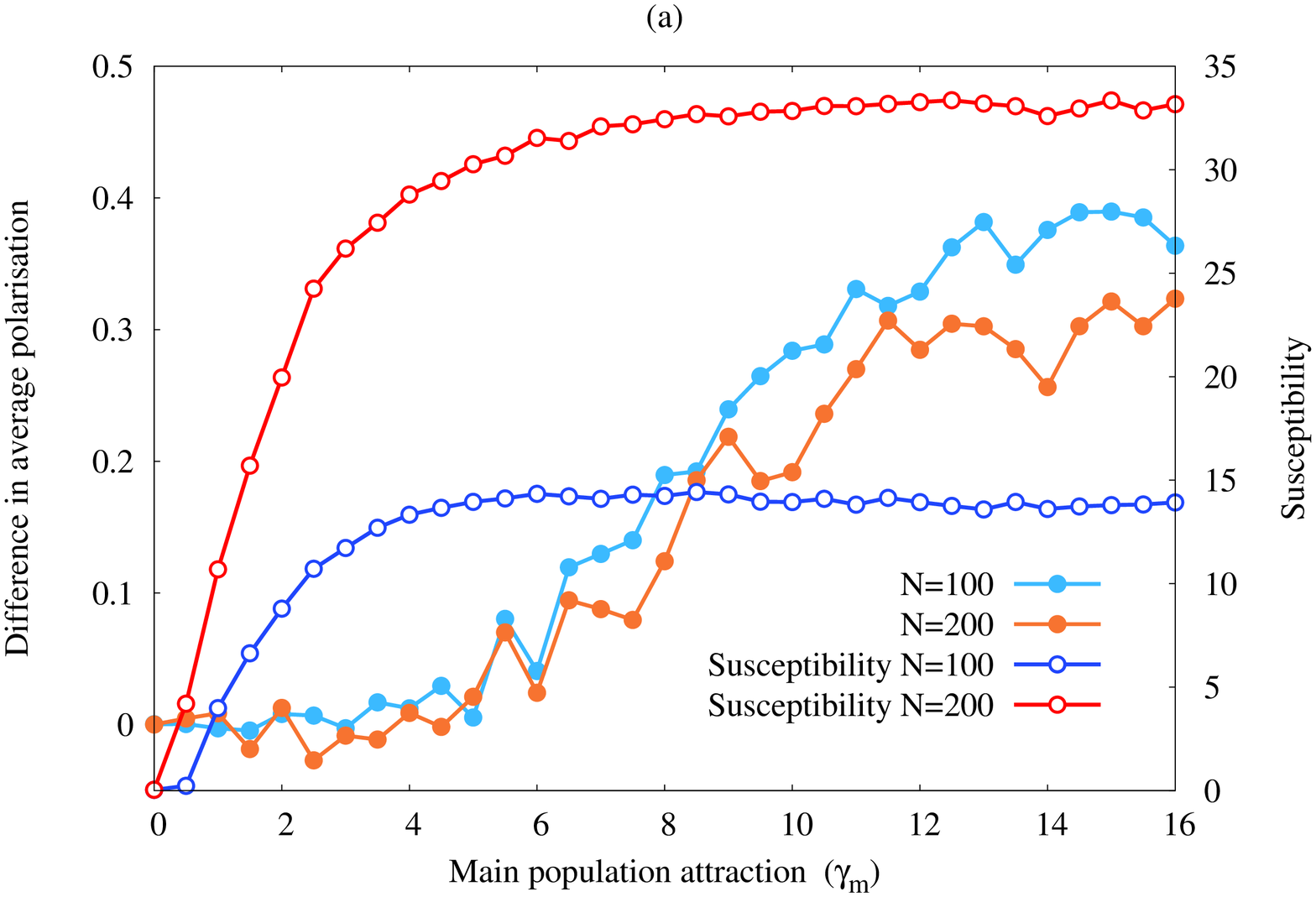}
 \caption{(a) Difference in average polarisation for 100 and 200 fish
   (light blue and orange filled circle respectively) with a single
   perturbing fish ($\gamma_p=14$ and $\beta_p=1$), along the
   transition region between schooling and milling defined by
   $\beta_m=A \sqrt{\gamma_m}+B$. $A$ and $B$ have been estimated for
   different group sizes: where $A=3.08$, $B=-1.97$ for $N=100$ and
   $A=3.28$, $B=-1.17$ for $N=200$. The corresponding susceptibility
   for unperturbed fish of 100 and 200 fish is shown on the right axis
   (blue and red hollow circle respectively).}
\label{PertTrans}
\end{center}
\end{figure}

Figure \ref{PertTrans} shows that even for a range of parameters where
the susceptibility has already reached a maximum value, the difference
in average polarisation still increases with the attraction parameter
$\gamma_m$ before it starts oscillating around the values shown at
$\gamma_m=16$. This means that while being in the transition region is
a required condition for a group of fish to exhibit sensitivity to
perturbations, a minimum level of attraction and alignment between
fish is required to significantly alter the group's response to these
perturbations. This additional requirement is probably due to the fact
that both $\gamma_m$ and $\beta_m$ increase while keeping a constant
noise, indicating that the main population only reacts to the
perturbation when the ratio noise to social interactions is kept below
a certain threshold.

\section{Discussion}

How interactions between individuals control the sensitivity to
perturbations of the group to which they belong and its ability to
respond to threats is an important issue to understand the evolution
of collective behaviours in animal swarms. The survival of each
individual within the group strongly depends on the capacity of
individuals to perform collective adaptive responses to different
conditions. Performing such responses not only requires coordination
mechanisms but also a high responsiveness to perturbations at the
group level, which can be favored by the presence of highly-correlated
fluctuations in the unperturbed state (illustrating again the deep
connection between response and fluctuation)
\cite{Attanasi2014b,Chate2014}.

Here we have addressed this question through an extensive
investigation of the responsiveness of a fish school model to
long-term standardised perturbations in the form of a single or a
small number of fish that display different interactions than the
main population in the school. We show that the school response
depends not only on the characteristics of the perturbing fish, but
of the collective state of the school as well. Indeed, in the
parameter space defining the way fish interacts with their
neighbours, there is a region that maximises the school response to
perturbations. This region is located throughout the transition
between schooling and milling states, where the school exhibits
multistability and regularly shifts between both states, and where
fluctuations are hence maximal.

The perturbing fish consists in agent(s) with intensities of the
attraction and alignment behaviours which differ from the rest of the
group. Borrowing the concept of susceptibility from magnetic systems
and other analyses of collective behaviour in biological systems
\cite{Cavagna2010, Attanasi2014}, we analysed its equivalent in our
simulations to measure the group's behavioural change caused by the
perturbing fish. We found that groups of fish display the highest
susceptibility in the transition region between the schooling and
milling states (figure \ref{Suscep}). If one assumes that our fish are
indeed in a transition region, the results presented here could be
compared to recent works \cite{Attanasi2014, Attanasi2014b,Bialek2014}
which have shown that animal swarms are in a critical state to better
adapt to various environmental conditions.

Indeed, the evolutionary advantages of social behaviour in animals can
easily be reduced if the organisms fail to adapt rapidly and/or
efficiently to a new challenging situation, for instance, in case of a
predator attack. It has been argued for some time that a more probable
solution for this problem is for a biological system to stay in a
perpetual state of transition from the most common behaviour
available, close to criticality \cite{Mora2011, Boedecker2012,
  Attanasi2014, Attanasi2014b,Bialek2014}, so that a minimal effort on
its part is able to push the collective behaviour into the new and
more appropriate one. Despite this, the system should also be
steady/robust enough to ignore certain perturbations and avoid
unnecessary transitions.

Our systematic study of the impact of perturbations in the parameter
space reveals that a group of fish in the transition zone is highly
affected by a perturbing fish with low attractions and/or alignment
values (figure \ref{PertScan}). When choosing which set of
parameters to use for the perturbing fish, we avoided the trivial
case where low attraction values coupled with the Voronoi
neighbourhood causes a following behaviour. For this reason, we used
a set of parameters that induced the perturbing fish to be located
closer to the group's centroid. We found that the group responds to
this perturbation by significantly increasing its level of
polarisation, shifting from a state in which the group spends half
the time in the schooling and milling states to a new state where
the group is schooling permanently (figures \ref{impact},
\ref{P100}(a) and \ref{P100}(b)).


We also checked whether the transition region is the only factor
involved in the group's responsiveness. As shown in figure
\ref{PertTrans}, even in the transition region (maximum
susceptibility), at lower values of $\gamma_m$ (and low $\beta_m$),
the school does not show the same responsiveness. The main
difference in this region compared to other transitions points is
the value of noise to social interactions ratio. 
This is in agreement
with the observed lower responsiveness of the school as the number
of perturbing fish is increased (insets of figures \ref{P100}(a) and
\ref{P100}(b)). This can be seen as an unnecessary large amount of
noise, which cancels the impact of the perturbation and decreases
the noise to social interactions ratio, as if the main population of
the school was located in lower values of the transition.

In this context, it is also interesting to mention the work of
Ioannou {\it et al}. \cite{Ioannou2012}, who studied the predatory
tactics of a fish towards a virtual school with different attraction
and alignment parameters. They found that the most frequently
attacked fish (figure 2 of \cite{Ioannou2012}) have low attraction
and/or low alignment parameters like the most influential perturbing
fish in our model (figure \ref{PertScan}). The fact that this
parameter region (low attraction and/or alignment) corresponds to a
vulnerability of the species (in Ioannou's work) and to a large
capacity to change the behaviour of the whole school, suggests that
when such deviant behaviour is detected it automatically triggers
the other group members to flee either from the current location
and/or from the vulnerable perturbing fish itself.

Previous works \cite{Romey1996, Kolpas2013} had already studied the
impact of perturbations in fish schools, but they focused on
punctual or instantaneous perturbations, while we looked into the
long-term changes that result from the perturbation. These analyses
were also mainly related to changes in the school trajectory, and
did not focus on the main behavioural changes undergone by the
school. When studying perturbations, one can either study how a
punctual change immediately affects the system, and the subsequent
recovering of the system to its original state, or one can
investigate what is the minimal constant perturbation imposed on the
system which is able to completely change its properties.

A similar approach to this continuous perturbation analysis has been
presented by Aureli {\it et al.} \cite{Aureli2010}, where they used a
self-propelled particle model to study the effect of an external
leader particle. However, Aureli {\it et al.} have chosen a perturbing
agent which is totally independent from the school's reactions. This
choice also enabled them to compare their results with experiments
performed with a remote controlled robotic fish able to influence a
school of Giant Danios ({\it Devario aequipinnatus})
\cite{Aureli2012}. Despite these similarities, there are two main
differences between the two models: (i) the total independence of the
perturbing agent, equivalent here to $\gamma_{p}=\beta_{p}= 0$; (ii)
the movement of the perturbing agent itself differs completely from
the school. More specifically, the perturbing agent and the school can
have different speeds. These differences limit the comparisons between
the two approaches. Nevertheless, for the case where the perturbing
agent has a similar speed as the rest of the particles, they observed
an increase in polarisation similar to the one we have found in
figures \ref{PertScan}, \ref{Pert_School}, and \ref{Pert_Mill}, for
the case $\gamma_{p}=\beta_{p}=0$.  Unfortunately, the observed state
in their simulations and experiments of agents/fish milling around the
perturbing agent cannot be reproduced here due to the fact that all
fish have the same speed. Given the model dependence on the fish
distance, in the case where the perturbing fish was fixed at some
point, the main population would be forced to remain close, either
swarming or milling around it, depending of the attraction parameters
$\beta_{m}$ used.

In order to best preserve the data-driven model developed by
Gautrais {\it et al}. \cite{Gautrais2009,Gautrais2012}, certain
limitations to the analysis arise. For instance, one can mention the
recent work by Couzin {\it et al.} \cite{Katz2011} which has shown
the importance of speed variation for the fish collective response.
However, in their original work, Gautrais {\it et al.}
\cite{Gautrais2009} observed that speed variation was minimal and
could be neglected for the considered species. Another limitation
results from the size of the school, given the dependence in fish
distance for their interactions. This term causes a limitation in
size of the school as seen in our previous work \cite{Calovi2014},
limiting simulations around the sizes of 100 and 200 fish presented
here.

In conclusion, our work has revealed that the collective states of a
school deeply influence its ability to respond to external or internal
perturbations. By providing a high responsiveness to perturbations,
the transition region between milling and schooling appears to be a
highly desired state that optimises the ability of the fish to react
collectively ({\it e.g.} to a predator attack), thus increasing the
survival of each individual within the school. Our results call for
further experimental observations on fish schools in order to measure
both their susceptibility and responsiveness to perturbations.

\section{Acknowledgements}

We are grateful to J.~Gautrais for inspiring discussions. We thank
the Laboratoire Plasma et Conversion d'Energie (CNRS UMR 5283) and
the Laboratoire de Physique Th\'eorique (CNRS UMR 5152) for providing
us  access to their computing facilities. D.\,S.\,C. was funded by
the Conselho Nacional de Desenvolvimento Cient{\'i}fico e
Tecnol\'ogico \^a Brazil. U.\,L. was supported by a doctoral fellowship
from the scientific council of the University Paul Sabatier. This
study was supported by grants from the Centre National de la
Recherche Scientifique and University Paul Sabatier (project
Dynabanc).



\newpage

\appendix{}

\newpage

\section{Supplementary figures}

\renewcommand{\figurename}{Supplemental Material, Figure}
\counterwithin{figure}{section}
\setcounter{figure}{0}

\begin{figure}[!htb]
\begin{center}
 \includegraphics[width=0.71\linewidth]{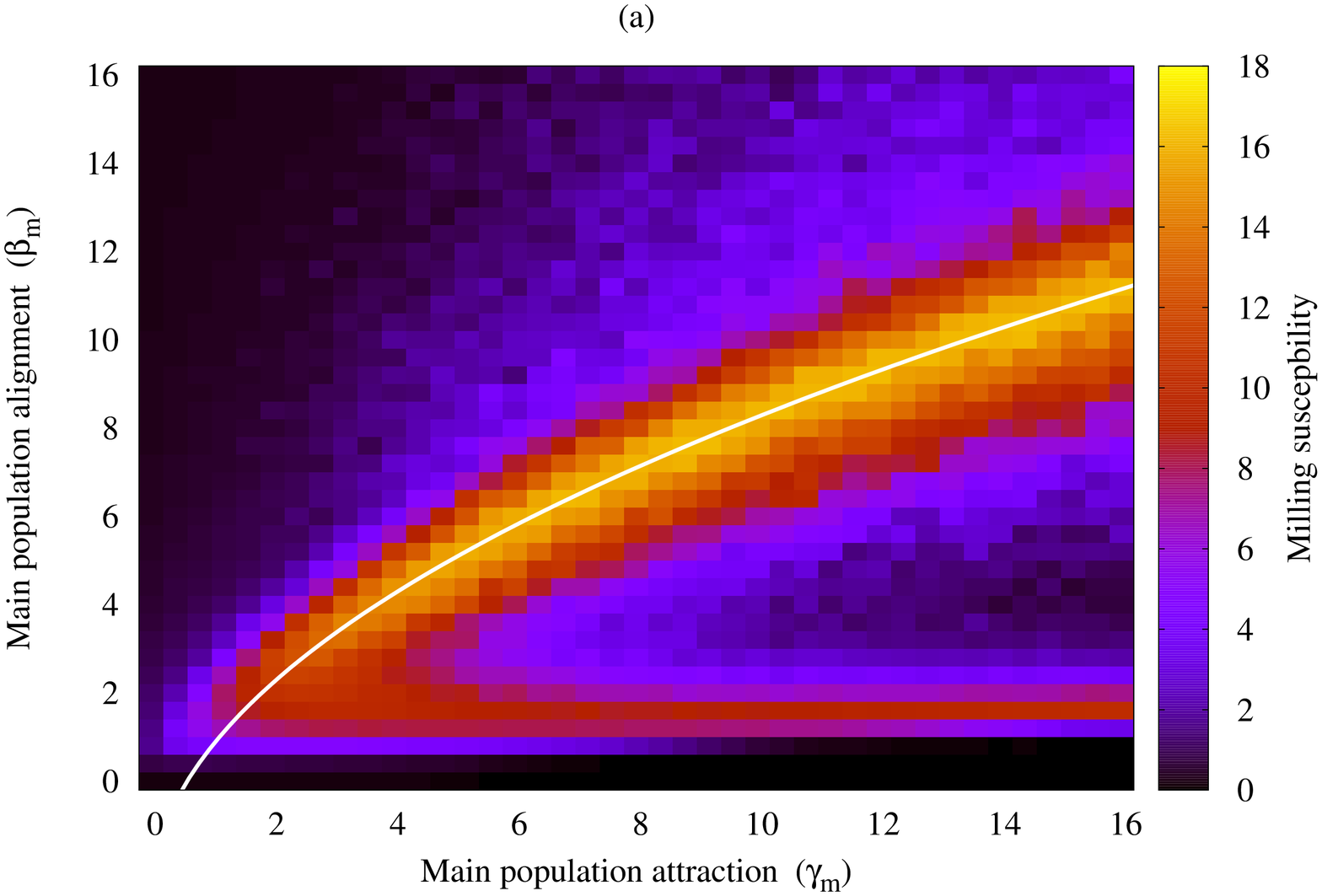}
 \includegraphics[width=0.71\linewidth]{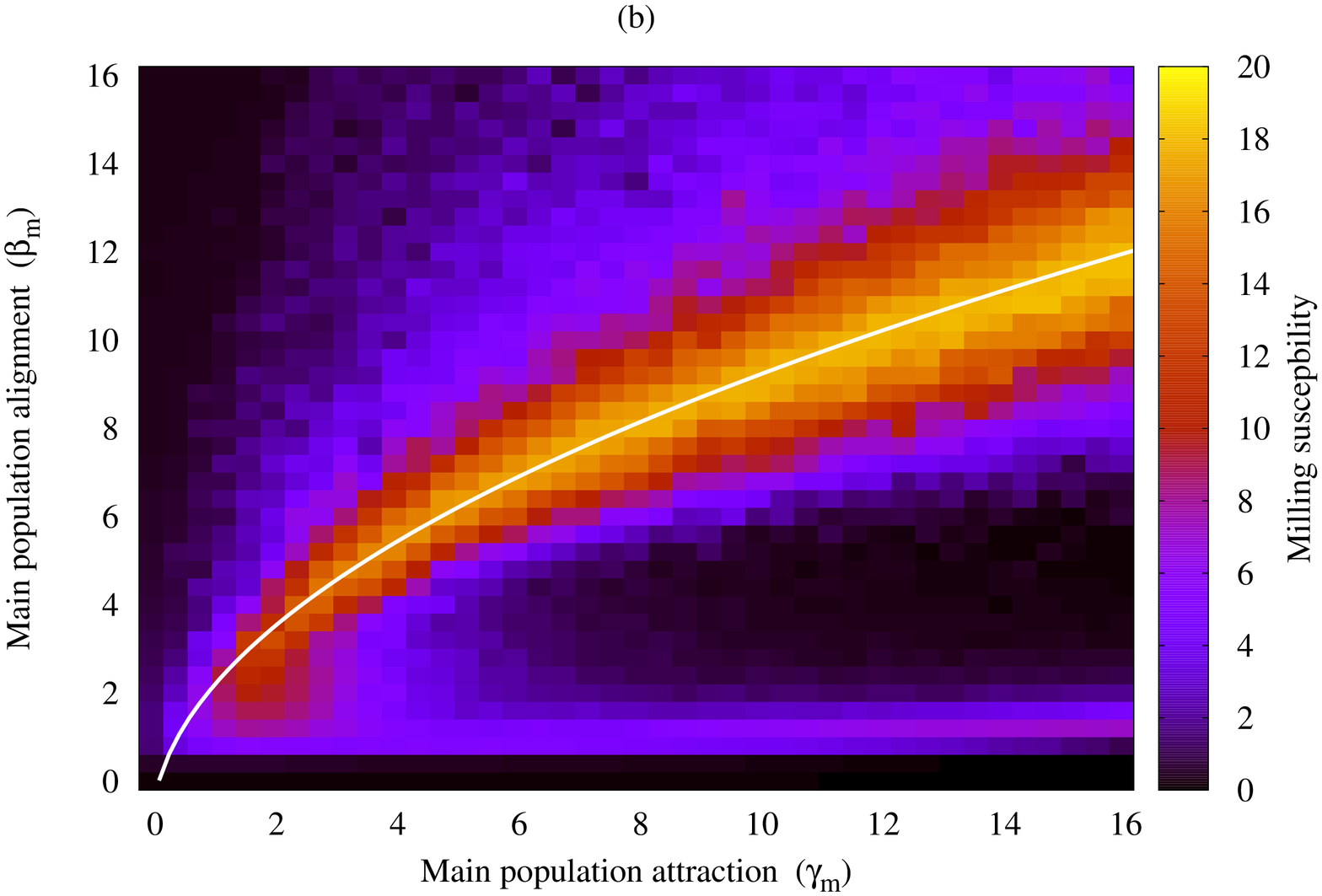}
 \caption{Milling susceptibility of unperturbed fish school
   simulations for 100 (a) and 200 (b) fish, as calculated from
   equation \ref{eqsuscm} (milling fluctuations), and for different
   values of the attraction and alignment parameters. Each data point
   represents an average over 400 simulations with random initial
   conditions. The white lines following the peak of susceptibility
   represents the function that fits the schooling/milling transition,
   as reported in \cite{Calovi2014}.}
\label{Suscep_Mill}
\end{center}
\end{figure}

In figure \ref{Suscep_Mill}, we present the milling susceptibility
$\chi_m$ measuring the fluctuations of the milling order parameter
in the unperturbed system, and introduced in equation \ref{eqsuscm}.

As already seen for the polarisation susceptibility $\chi$ (see
figure \ref{Suscep} in the main text), the milling susceptibility
$\chi_m$ should sharply increase near the schooling-milling
transition line, but should also be more sensitive to the narrow
winding phase than the polarisation susceptibility, near the
winding-milling transition line. Indeed, this is confirmed in figure
\ref{Suscep_Mill}(a) where the schooling-milling transition is
clearly identified (as it was by using the polarisation
susceptibility), and where the narrow winding-milling transition
line is much more clearly apparent than by using the polarisation
susceptibility (compare figure \ref{Suscep}(a) to figure
\ref{Suscep_Mill}(a)). In addition, the weakening of the milling
susceptibility enhancement near the winding-milling transition as
one increases the number of fish $N$ observed in figure
\ref{Suscep_Mill}(b) (as compared to figure \ref{Suscep_Mill}(a))
strongly suggests that the winding phase probably disappears for
larger $N$, its very elongated shape making it more unstable as $N$
increases.

\begin{figure}[!htb]
\begin{center}
 \includegraphics[width=0.7\linewidth]{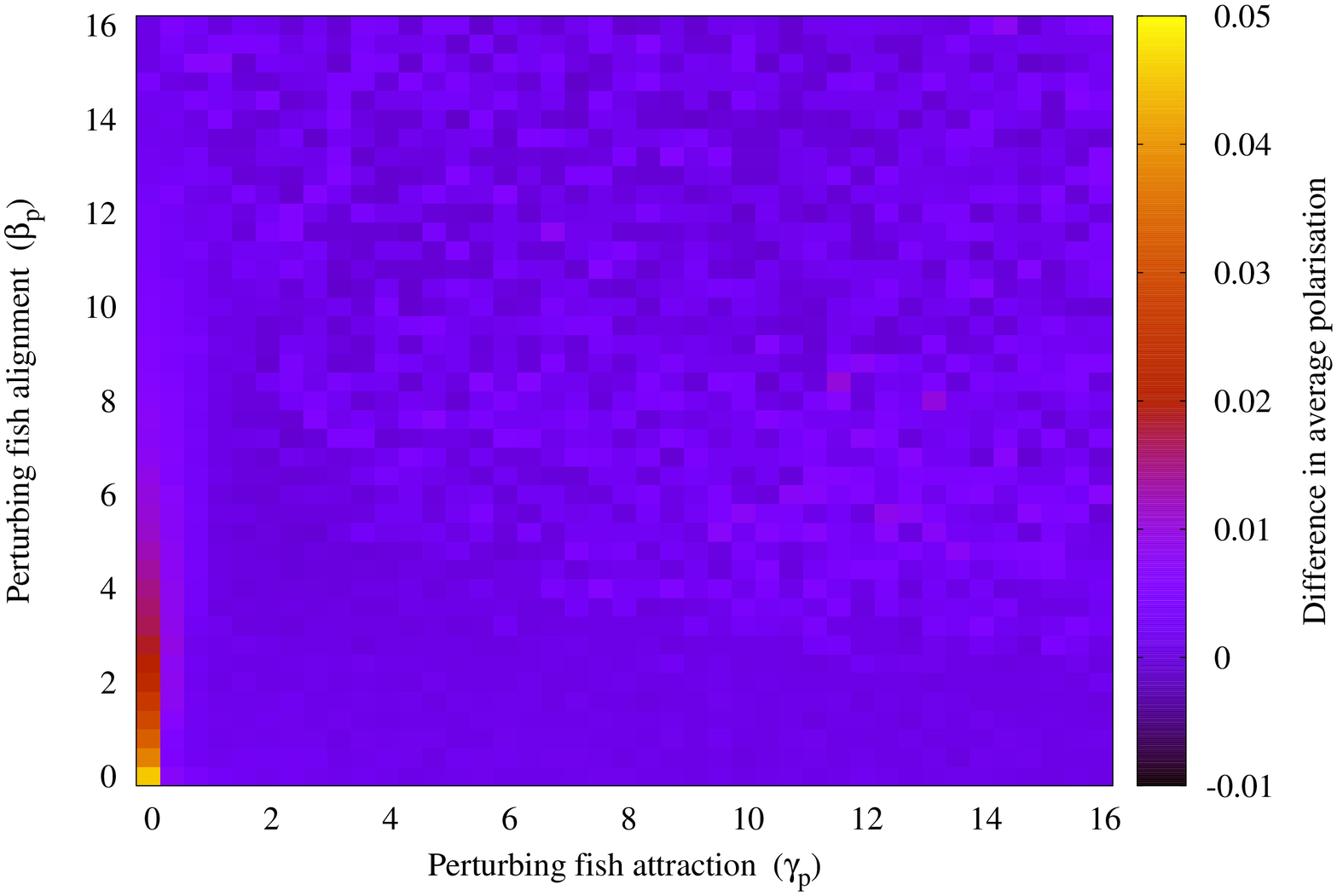}
\end{center}
 \caption{Each data point represents the difference in the average
   polarisation caused by a single perturbing fish, while varying its
   attraction ($\gamma_p$) and alignment ($\beta_p$) parameter, in
   comparison to the unperturbed case located on the schooling region
   ($\gamma_m= 4$, $\beta_m= 14$).}
\label{Pert_School}
\end{figure}

\begin{figure}[!htb]
\begin{center}
 \includegraphics[width=0.67\linewidth]{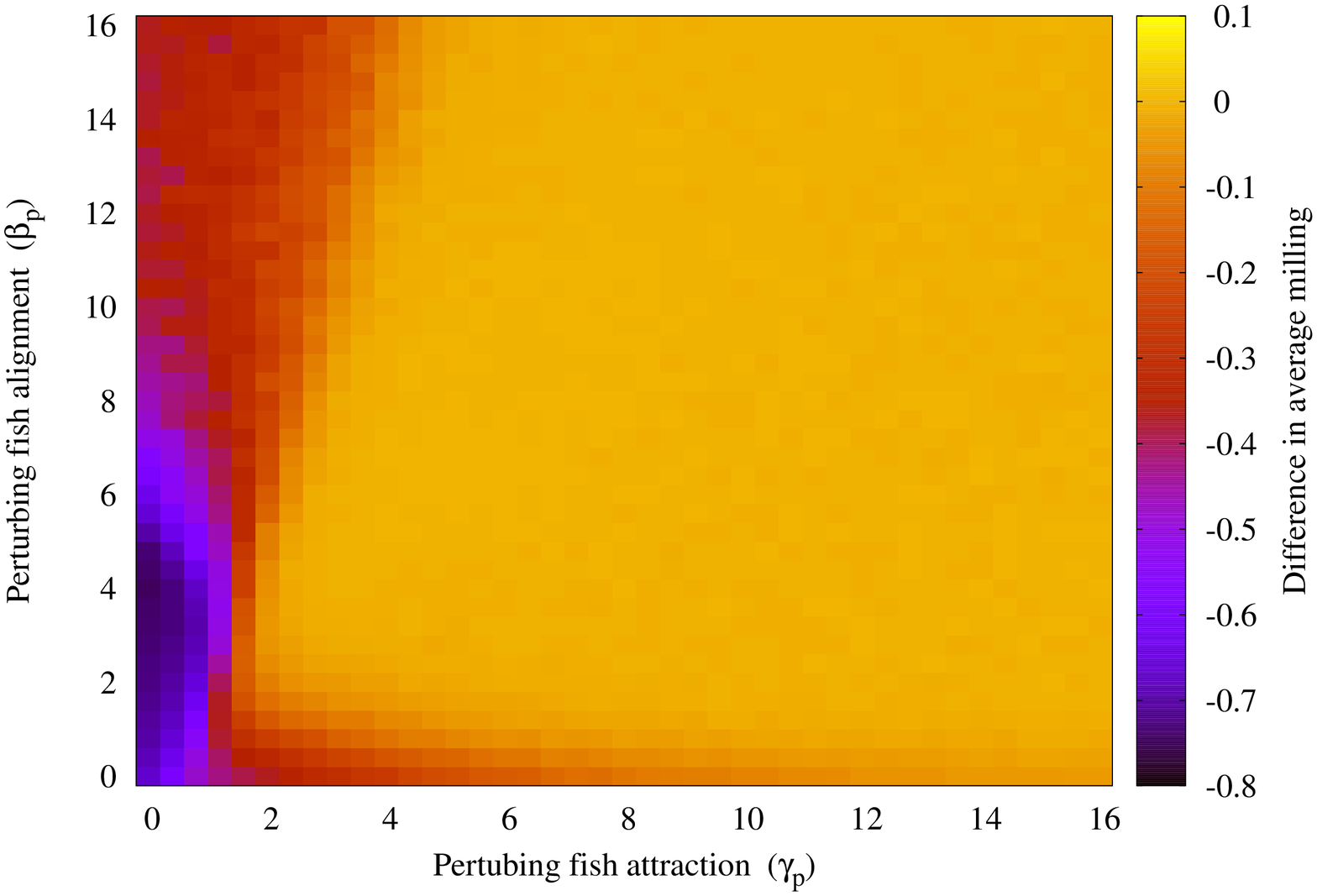}
\caption{Each data point represents the difference in the average
  polarisation caused by a single perturbing fish, while varying its
  attraction ($\gamma_p$) and alignment ($\beta_p$) parameter, in
  comparison to the unperturbed case located on the milling region
  ($\gamma_m=14$, $\beta_m= 4$).}
\label{Pert_Mill}
\end{center}
\end{figure}

\newpage
\section{Supplementary movies}

\renewcommand{\figurename}{Supplemental Material, Movie}
\counterwithin{figure}{section}
\setcounter{figure}{0}

\begin{figure}[!htb]
\caption{Inset3.mp4 - Simulation with 100 fish where $N_p=1$ (red
  fish), $\gamma_p=1$, $\beta_p=7$ and the main population located in
  the transition between schooling and milling ($\gamma_m=14$,
  $\beta_m=10$).}
\label{inset3}
\end{figure}

\begin{figure}[!htb]
\caption{Inset4.mp4 - Simulation with 100 fish where $N_p=1$ (red
  fish), $\gamma_p=1$, $\beta_p=1$ and the main population located in
  the transition between schooling and milling ($\gamma_m=14$,
  $\beta_m=10$).}
\label{inset4}
\end{figure}

\begin{figure}[!htb]
\caption{Inset6.mp4 - Simulation with 100 fish where $N_p=1$ (red
  fish), $\gamma_p=14$, $\beta_p=1$ and the main population located in
  the transition between schooling and milling ($\gamma_m=14$,
  $\beta_m=10$).}
\label{inset6}
\end{figure}


\end{document}